\documentstyle[floats,multicol,aps,epsf,prb]{revtex}

\begin{document}
\ifx\href\undefined\else\errmessage{hyperTeX disabled by xxx admin}\fi
\draft
\twocolumn[\hsize\textwidth\columnwidth\hsize\csname @twocolumnfalse\endcsname
\title{Ultrasound attenuation in gap-anisotropic
systems}
\author{J. Moreno and P. Coleman}
\address{Department of Physics and Astronomy, Rutgers University,
Piscataway, NJ 08855}
\date{\today}
\maketitle

\begin{abstract}
Transverse ultrasound attenuation provides a weakly-coupled probe of
momentum current correlations in electronic systems.  We develop a
simple theory for the interpretation of transverse ultrasound
attenuation coefficients in systems with nodal gap anisotropy.
Applying this theory we show how ultrasound can delineate between
extended-s and d-wave scenarios for the cuprate superconductors.
\end{abstract}
\vskip2pc]
\input psfig.tex

The development of gaps with nodal anisotropy is a recurrent property
of highly correlated electron systems.  Bulk probes, such as the NMR
relaxation rate,\cite{kitaoka} specific heat\cite{steglich} and
penetration depth \cite{broholm,Hardy} indicate that gap
nodes may be present in a wide variety of strongly correlated
systems including heavy fermions, strong coupling  and cuprate superconductors
\cite{hfermion,v3si,cuprates}
and the narrow gap kondo  insulators
$CeNiSn$ and $CeRhSb$.\cite{takabatake}
However, with a few noted
exceptions\cite{upt3review,kirtleyreview}, 
we have no direct information about the symmetry
of the gap in these strongly correlated systems.

A versatile, but much under-utilized tool for probing electronic gap
nodes is the use of transverse ultrasound attenuation.  This method has
been successfully used to locate the gap lines and point nodes in
superconducting $UPt_3$.\cite{shivaram} Surprisingly, very little
work has been done to enable the model-independent
interpretation of transverse ultrasound measurements.
In this paper we revisit this old problem,
highlighting those  aspects of ultrasound attenuation that are
model-independent and relevant to  future experiments.

Ultrasound attenuation probes the relaxation of electronic momentum in
a model-independent fashion.  This information is encoded in the
``viscosity tensor'', a high symmetry tensor with very few
independent components. Here  we develop a simple theory which
links these components to the 
location of the gap nodes. We illustrate this theory in a vein
of current interest, cuprate
superconductivity, showing how ultrasound measurements can provide an
unambiguous fingerprint of gap zeroes lying on the diagonal of the
Brillouin zone.\cite{ultatt}

In a metal, the phonon
strain field $u_{ij}(x)$ couples linearly to the electron stress tensor 
$\sigma^{ij}(x)$ 
\begin{equation}
H_I = - \int d^3 x \sigma^{ij}(x) u_{ij}(x).
\end{equation}
This coupling is model-independent.  
The stress tensor $\sigma^{ij}(x)$ describes the flow of
electronic momentum: its divergence governs the
rate of change of electronic momentum density $\nabla_j \sigma^{ij}(x)
= -\dot P_i(x)$.  
When a sound
wave propagates through a crystal, the dissipation rate is:
\cite{thermal}
\begin{equation}
\dot {E} = - \int d^3 x \sigma^{ij}(x) {\dot{u}}_{ij}
\end{equation}
In linear response
$\sigma^{ij}= \eta^{ijkl} {\dot{u}}_{kl}$ where $\eta^{ijkl}$
is the viscosity tensor.  The sound attenuation coefficient is defined
as the ratio of the time average energy 
dissipation to twice the energy flux in the wave:
\cite{Rodriguez}
\begin{equation}
\alpha(\vec{q},\hat{u})=\frac{q^2}{\rho c_s} \overline{\eta}
\hspace{0.2in}{\rm where} \hspace{0.15in} 
\overline{\eta} = \eta^{ijkl}\hat{u_i}\hat{q_j}\hat{q_k}\hat{u_l},
\end{equation}
$\rho$ is the mass density and $c_s$ is the speed of 
sound with wave vector 
$\vec{q}$ and polarization $\hat{u}$.
From the coupling (1), 
it is straight-forward to obtain
the Kubo formula for the
viscosity tensor:
\begin{equation}
\eta^{ijkl}(\vec{q}) = -  \hspace{0.05in} lim_{\omega \rightarrow 0} 
\frac{1}{i \omega}
\langle  \sigma^{ij}
({\bf q})\sigma^{kl} (-{\bf q}) \rangle 
\end{equation}
with  ${\bf q}=(\vec{q},\omega + i\delta)$.

In typical ultrasonic measurements, wavelengths $\lambda$
are  hundreds of  microns and 
substantially  exceed the electronic
mean-free paths ($l_e$), so the attenuation is safely in the
hydrodynamic limit, $l_e<<\lambda$. 
In this case the momentum dependence of the 
viscosity tensor is irrelevant, permitting us to take the limit 
$\vec{q} \rightarrow 0$.\cite{ultatt}

Like the elasticity tensor, 
the symmetry properties of the viscosity tensor 
significantly reduce  the number of its independent components.
\cite{Landau}
This tensor is symmetric, not only
in the first and second pair of indices, but also under the
interchange of these pairs:
\begin{equation}
\eta^{iklm}=\eta^{kilm}=\eta^{ikml}=\eta^{lmik}
\end{equation}
Symmetry under the crystal point group
further reduces the number of independent
components.
For example,  inversion symmetry 
eliminates those components with an odd number of identical suffixes,
e.g. $\eta^{xxxy}=0$, $\eta^{xyyy}=0$.
In a square or cubic environment, $90^0$ rotation symmetry
restricts the viscosity tensor to the form
\begin{equation}
\eta^{ijkl}= A(\delta^{ik}\delta^{jl} + \delta^{il}\delta^{jk})
+\tilde{C}\delta^{ij}\delta^{kj} + B \delta^{ij}\delta^{jk}\delta^{kl}
\end{equation}
with no sum on indices implied, 
so for a cubic or
square lattice the ultrasound attenuation is proportional to:
\begin{equation}
\overline{\eta}= A + 
B \sum_{i=1,d} \hat u_i^2 \hat q_i^2+
C(\hat u \cdot \hat q)^2,
\end{equation}
where $C=\tilde{C}+A$ and $d$ is the dimension.
We shall restrict our attention to transverse ultrasound
attenuation, for which the last term vanishes. In other words,
in a cubic crystalline environment, there are only two
independent transverse ultrasound attenuation coefficients.
In a  two dimensional  hexagonal system,
$60^0$ rotation symmetry means that $B$ must also
vanish.
In a $3D$ hexagonal system, polynomial terms 
involving $\hat u_z$ and $\hat q_z$ must be added to the
above expression, giving
\begin{equation}
\bar \eta= A + 
B(\hat{u}_{z}^2 + \hat{q}_z^2) + C \hat{u}_{z}^2
\hat{q}_z^2 
\end{equation}
where we have dropped terms which vanish for transverse ultrasound.
These  strong symmetry constraints mean that
only a few different propagation directions are required to measure
the full electronic viscosity tensor.

We now turn to the relationship between the quasiparticle gap structure
and the
electron viscosity tensor.
The electron contribution to ultrasonic
attenuation only becomes substantial at low temperatures in a
regime where a quasiparticle description of the excitations is valid.
For a fluid of quasiparticles with dispersion $E_{\vec k}$, the
quasiparticle group velocity is $\vec v_{\vec k} = \nabla_{\vec k}
E_{\vec k}$.  The 
traceless stress tensor pertinent to  transverse
ultrasonic measurements is then:
\begin{equation}
\sigma^{ij} = \sum_{\vec k \sigma}\psi^{\dagger}_{\vec k \sigma}
\sigma^{ij}_{\vec k}
\psi_{\vec k \sigma},
\end{equation}
where $\psi^{\dagger}_{\vec k \sigma}$ creates a quasiparticle
of momentum $\vec k$ and 
\begin{equation}
\sigma^{ij}_{\vec k}
={1 \over 2}(k^i v_{\vec k}^j+
k^j v_{\vec k}^i)- {1 \over d}
\delta^{ij}\vec{k}\cdot \vec{v}_{\vec k}
\end{equation}
is the momentum flux of a single quasiparticle.  The
quasiparticle contribution to the viscosity tensor is then
\begin{equation}
\eta^{ijkl}=\sum_{\vec{k}} (-\frac{\partial{f}}
{\partial{E_k}} ) \tau_{\vec{k}} \sigma^{ij}_{\vec k}\sigma^{kl}_{\vec k},
\end{equation}
where $f$ is the Fermi distribution function and
$\tau_{\vec{k}}$ is the
relaxation time of the quasiparticle.
From this relation, we see that a simple expression for the
viscosity for transverse ultrasound is:
\begin{equation}
\bar \eta ={1 \over 4} \sum_{\vec{k}} \left(-
{\partial{f}\over 
\partial{E_k}} \right) \tau_{\vec{k}} 
\biggl([\vec v_{\vec k}\cdot \hat u][\vec k\cdot \hat q]
+
[\vec k\cdot \hat u][\vec v_{\vec k}\cdot \hat q] 
\biggr)^2
\end{equation}

We shall consider the situation where the temperature is low
enough for the quasiparticles to be entirely concentrated within
the gap nodes of the excitation spectrum.  
To simplify our discussion,
we shall assume that the gaps are small enough in comparison with
the Fermi energy, that to a good approximation
$\vec v_{\vec p}  \approx v_F \hat p_F\left(
\partial  E_{\vec p}/\partial \epsilon_{\vec p}\right)$,
where $\epsilon_{\vec p}$ and $v_F$ are respectively, the 
``bare'' energy and Fermi velocity of the
excitations prior to gap formation.
Consider the case where the gap nodes
are simple points in momentum space, located at positions
$\vec p_o(i)$. 
The attenuation from a given node
will depend on the orientation of the sound wavevector and polarization.
If neither of these vectors is perpendicular to $\vec p_o(i)$ then 
the quasiparticles at the node can couple to
the sound-wave. In this configuration, the node
is {\em ``activated''}. (Fig. 1.a)
If however, either the
wavevector direction $\hat q$ or the polarization $\hat u$ are
perpendicular to the node, it is {\em ``inactive''} and
quasiparticles at the bottom
of the node will not couple to the ultrasound. (Fig. 1.b) In this 
configuration, the
attenuation produced by the node is strongly suppressed.
For most orientations of the ultrasound, the nodes are ``activated'', and
their contribution to the ultrasound attenuation may be written as:
\begin{figure}[btp]
\psfig{file=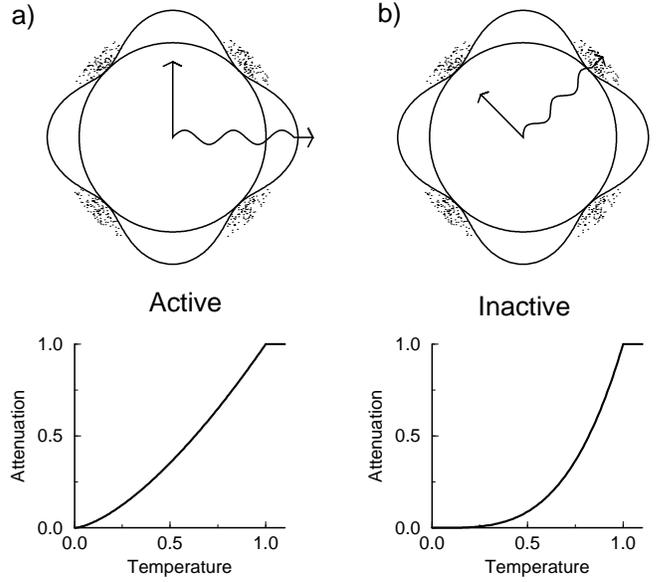,width=3.375in}
\caption{a) The nodes are ``activated'', b) $\hat{q}$ or $\hat {u}$
are perpendicular to the nodes, the nodes are ``inactive''}
\label{fig1}
\end{figure}
\begin{equation}
\bar
\eta_{A}
=  (v_F p_F)^2 \sum_{i} ( \hat u \cdot \hat p_o(i))^2 (\hat q \cdot 
\hat p_o(i) )^2 {\cal A}_i (T),
\end{equation}
where 
\begin{eqnarray}
{\cal A}_i (T) &=&
\overline{
(2 N_i(E) \Gamma_i(E))^{-1} } \nonumber \\
&=& \int_{\vert\vec p- \vec p_o(i)\vert < \Lambda}
{d^dp \over (2 \pi)^d }
\left(-
{\partial f\over \partial E_{\vec p}
}\right)
\left(
{\partial E_{\vec p} \over \partial \epsilon_{\vec p}
}\right)^2 \tau_{\vec p}
\end{eqnarray}
is the thermal average of the inverse product of twice the 
relaxation rate $\Gamma_{i}(E)$ and the quasiparticle
density of states $N_{i}(E)$ in the  gap node.

Suppose instead the gap node is ``inactive'', with the polarization
vector $\hat u$ at right-angles to $\vec p_o(i)$, then the contribution
to the ultrasonic attenuation contains the additional factor
$\cos^2 (\theta)$, where $\cos \theta = (\hat k \cdot \hat u) $. 
In this case:
\begin{equation}
\bar
\eta_{I}(i)
=  (v_F p_F)^2 
(\hat q \cdot 
\hat p_o(i) )^2 {\cal B}_i(T),
\end{equation}
where 
\begin{eqnarray}
{\cal B}_i (T) =
\overline{
\cos^2 \theta_{\vec p}(2  N_i(E) 
\Gamma_i(E))^{-1} }= 
\nonumber \\
=\int_{\vert\vec p- \vec p_o(i)\vert < \Lambda}
{d^dp \over (2 \pi)^d }
\left(
-
{\partial f\over \partial E_{\vec p}
}\right)
\left(
{\partial E_{\vec p} \over \partial \epsilon_{\vec p}
}\right)^2 \tau_{\vec p}\cos^2 {\theta^2_{\vec p}}
\end{eqnarray}
For a point-node where the size of the node grows
linearly with energy, ${\cal B}_i(T) \sim T^ 2 
{\cal A}_i(T)$.  
A similar result holds for a line node which lies
in a plane. 
It thus follows that if a configuration can be found
where all nodes are simultaneously inactive, then
the ultrasonic attenuation will exhibit a relaxation
rate a factor of $T^2$ smaller.
Such a situation will occur for point nodes situated
at $90^0$ to each-other.  It will also occur for  line
nodes lying in a plane,
or lying  in many planes that intersect
at $90^0$.  If the nodes do not lie
in such $90^0$ configurations, then ultrasound attenuation
will not show an anisotropic dependence of
the powerlaws.

To illustrate this discussion, we now make a more
detailed application to a two-dimensional model
relevant to the cuprate superconductors.
A number of recent experiments have provided strong evidence for an
anisotropic gap with nodes at the Fermi surface
\cite{Hardy,Ishida}. 
Superconducting interference experiments
sensitive to the phase of the gap function
\cite{Wollman} 
support an order parameter with d-wave symmetry with
nodes lying at $45^0$ to the a and b axis.
Ultrasound measurements provide a complimentary approach
which is sensitive to the precise location of the nodes.
In particular, it offers the potential to 
distinguish between a $d_{x^2-y^2}$ state and an
anisotropic s-wave state with nodes located either side
of the $45^0$ position, and there are no ambiguities
associated with the interpretation of results for bilayer compounds.

Using the symmetry arguments advanced above, 
if the wave vector $\vec{q}$ forms an angle $\phi$
with the x-axis, the transverse  ultrasound attenuation
$\alpha^{T}(\phi)$,  will have the form 
\begin{eqnarray}
\alpha^{T}(\phi)= [\alpha^{T}(\pi/4)- \alpha^{T}(0)] \sin^{2}2\phi +
\alpha^{T}(0) ; 
\end{eqnarray}
Suppose the quasiparticle excitation spectrum has the BCS 
form $E_{\vec p} = \sqrt{\epsilon_{\vec p}^2 + \Delta_{\vec p}^2 }$.
Using the results obtained above, this leads to 
\begin{eqnarray}
\frac{\alpha(\vec{q},\hat{u})}{\alpha_N(\vec{q},\hat{u})}
= \frac{1}{<\Pi^2(\vec{q},\hat{u})>}
\int  d\omega (-\frac{\partial{f}}{\partial{\omega}})
\frac{ \tau(\omega)}{\tau_N} & &\nonumber \\
\int_{0}^{2\pi} \frac{d\theta_p}{2\pi}
Re(\frac{\sqrt{\omega^2-|\Delta_p|^2}}{\omega}) \Pi^2(\vec{q},\hat{u})
& &.
\label{1}
\end{eqnarray}
where,  $\Pi(\vec{q},\hat{u})=
(\hat{p}\cdot\hat{u})(\hat{p}\cdot\hat{q})$ and
$\alpha_N$ and $\tau_N$ are the normal attenuation and
relaxation times. 

To represent the d-wave order parameter we have chosen the function:
\begin{equation}
\Delta_{d} (\theta_p)=\Delta_{d} \cos{2\theta_p}
\end{equation}
 The anisotropic s-wave is
represented by: 
\begin{equation}
\Delta_{s} (\theta_p)=
\Delta_{s} \Phi_{s}(\theta_p) + \langle \Delta \rangle
\end{equation} 
where:  $\Phi_{s}(\theta_p)=|\cos(2\theta_p)| -(2/\pi)$.
We have treated impurity scattering in the standard 
selfconsistent T-matrix approximation 
\cite{Schmitt}.

\begin{figure}[btp]
\psfig{file=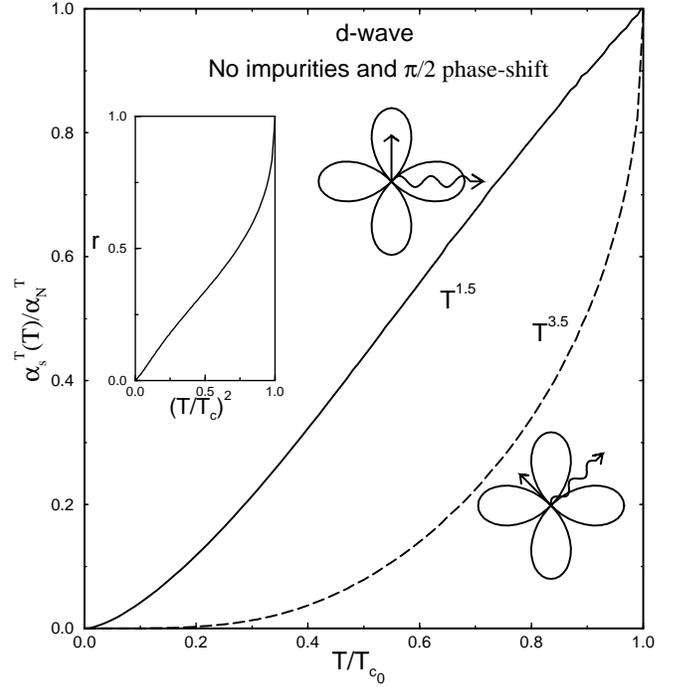,width=3.375in}
\caption{Ultrasound attenuation for a d-wave state at two angles:
$\phi=0$  (solid line) and $\phi=\pi/4$ (dashed line). In the inset
the ratio $r=\alpha_{s}(\pi/4)/\alpha_{s}(0)$  as a
function of $(T/T_{c})^2$.}
\end{figure}

For a pure d-wave state (Fig. 2) the transverse ultrasound 
attenuation is proportional to
$T^{1.5}$ when the angle is zero (the wave vector and polarization of the
sound wave are parallel to the symmetry axis of the crystal) and it is
proportional to $T^{3.5}$ when the angle is $\pi/4$.
When there is a finite density of impurities a flat region appears in
both attenuation coefficients (the impurities induced a finite density
of quasiparticle states at zero temperature), but at higher
temperatures the power laws are unchanged.
Then, in accord with our earlier arguments, 
\begin{eqnarray}
\frac{\alpha^{T}(\pi/4)}{\alpha^{T}(0)} \propto 
\frac{T^{3.5}}{T^{1.5}}= T^2
\end{eqnarray} 
Even considering non-resonant scattering (taking the cotangent of the
scattering phase shift c=1) this ratio shows a quadratic behavior.

The results for a pure s-wave state are very different (Fig.3). 
We have chosen
two different anisotropic s-waves, $\langle \Delta \rangle=\frac
{2}{\pi} \Delta_s$ and $\langle \Delta \rangle=0$.
Unlike d-wave pairing, here the anomalous scattering off impurities
is finite and resonant scattering does not ever develop.
 \cite{Hirschfeld Stanford}. Consequently,
the attenuation for an  s-wave gap with nodes at $45^0$ is even more
anisotropic, it is finite when the angle is zero and approaches zero
when $\phi=\pi/4$.
On the other hand, when the
nodes of the gap function depart from $\pi/4$ the ultrasound
attenuation anisotropy disappears, in fact at temperatures close enough to
$T_c$, the attenuation at $\phi=\pi/4$ actually becomes bigger
that at $\phi=0$. {\em In neither case} is
the ratio  $\alpha^{T}(\pi/4)/\alpha^{T}(0)$ 
proportional to $T^2$.

\begin{figure}[btp]
\psfig{file=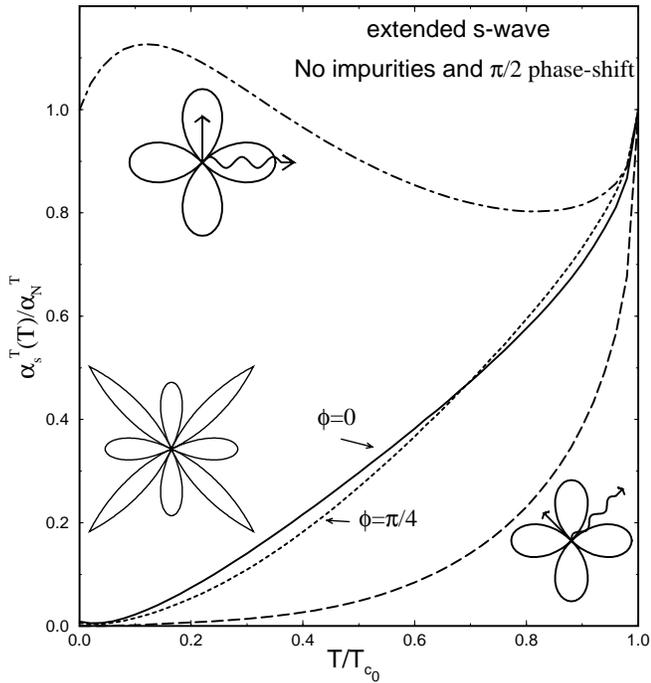,width=3.375in}
\caption{Ultrasound attenuation for two extended s-wave states: 
$\Delta_{s} (\theta_p)=\Delta_{0}| \cos{2\theta_p}|$ at angles
$\phi=0$  (dot-dashed line) and $\phi=\pi/4$ (dashed line), and
$\Delta_{s} (\theta_p)=\Delta_{0}(| \cos{2\theta_p}|-(2/\pi))$ at 
$\phi=0$  (solid line) and $\phi=\pi/4$ (dotted line).} 
\end{figure}

The generality of our approach lends itself naturally
to many other strongly correlated 
systems with point nodes.  One particularly interesting case is the narrow
gap Kondo insulators, $CeNiSn$ and $CeRhSn$, where Miyake et al 
\cite{Miyake}
have suggested point-nodes on the c-axis as an explanation of the
unusual NMR relaxation rate and the anisotropy of the conductivity.
There are several other strongly
correlated superconductors that deserve examination, such
as $V_3Si$ and $UPd_2Al_3$ which exhibit $T^3$ NMR
relaxation rates characteristic of gap lines.
The heavy fermion
superconductor $UBe_{13}$ is noteworthy here, early ultrasound
measurements \cite{Batlogg}, found
no anisotropy, despite the clear suggestion of line nodes from NMR
measurements.\cite{maclaughlin} This result suggests the presence 
of domains
with different orientations, and it would 
be interesting to repeat the measurements using field-cooling
to obtain an aligned single domain superconductor.  

This paper has emphasized the model-independent aspects of ultrasound
attenuation as a probe of nodal gap structure. 
This attenuation is independent of the
mechanism driving superconductivity and sensitive only to the
intrinsic symmetries of quasiparticle excitation spectrum. Its
tensorial character permits 
the measurement of several independent components at the same time. 
The simple methods developed provide an economic way
to extract vital information about the gap anisotropy in a general
class of gap-anisotropic systems, and they appear to 
provide a discriminating tool for elucidation of gap structure in the cuprates.

We are grateful to S. Bhattacharya for useful discussions. This
research was supported by NSF grant DMR-93-12138.


\begin{references}
\bibitem{thermal} We have neglected a contribution due to the
electronic thermal conduction.
This term appears only for longitudinal
attenuation and it is small compared to the viscosity term.
\end{references}
\end{document}